\documentclass[12pt]{article}

\usepackage[super]{natbib} 
\usepackage{color} 
\usepackage[table]{xcolor} 
\usepackage{paralist}
\usepackage{geometry}
\usepackage{graphicx,graphics}
\usepackage{amsmath}	
\usepackage{amssymb}	
\usepackage{newtxtext,newtxmath}
\usepackage[T1]{fontenc}
\usepackage{ae,aecompl}
\usepackage{epsfig}   
\usepackage{microtype}
\usepackage[utf8]{inputenc}
\usepackage{enumitem}
\usepackage{ragged2e}
\usepackage{times}
\newlist{thematic}{itemize}{8}
\setlist[thematic]{label=$\square$}
\usepackage{pifont}
\usepackage{enumitem} 
\usepackage{hyperref} 
\hypersetup{
    colorlinks,
    linkcolor={blue!80!black},
    citecolor={blue!80!black},
    urlcolor={blue!80!black}
}

\def\araa{{ARA\&A}}          
\def\apj{{ApJ}}                 
\def\apjl{{ApJ}}

\def\aap{ {A\&A}}                
\def\aapr{ {A\&A~Rev.}}

\def\mnras{ {MNRAS}}

\def\pasa{ {Publ.~Astr.~Soc.~Australia}}

\newcommand{\be}{\begin{equation}}
\newcommand{\ee}{\end{equation}}

\newcommand{\msun}{{$M_{\odot}$}}
\newcommand{\mstar}{{$M_{\star}$}}

\newcommand{\gtsima}{$\; \buildrel > \over \sim \;$}
\newcommand{\ltsima}{$\; \buildrel < \over \sim \;$}
\newcommand{\prosima}{$\; \buildrel \propto \over \sim \;$}
\newcommand{\gsim}{\lower.5ex\hbox{\gtsima}}
\newcommand{\lsim}{\lower.5ex\hbox{\ltsima}}
\newcommand{\simgt}{\lower.5ex\hbox{\gtsima}}
\newcommand{\simlt}{\lower.5ex\hbox{\ltsima}}
\newcommand{\simpr}{\lower.5ex\hbox{\prosima}}

\newcommand{\cxo}{\textit{Chandra}}

\newcommand{\mbh}{$M_{\rm BH}$}
\newcommand{\lx}{$L_{\rm X}$}

\newcommand{\Chandra}{\textit{Chandra}}

\newcommand{\factive}{$f_{\text{active}}$}
\newcommand{\focc}{$f_{\text{occ}}$}

\newcommand{\lxrb}{$L_{X,\text{XRB}}$}
\newcommand{\pxrb}{$P_{\text{XRB}}$}

\begin{document}

\pagenumbering{roman}

\raggedright
\huge
Astro2020 Science White Paper \linebreak

Towards a high accuracy measurement of the local black hole occupation fraction in low mass galaxies \linebreak
\normalsize

\noindent \textbf{Thematic Areas:} \hspace*{60pt} $\square$ Planetary Systems \hspace*{10pt} $\square$ Star and Planet Formation \hspace*{20pt}\linebreak
$\boxtimes$ Formation and Evolution of Compact Objects \hspace*{31pt} $\square$ Cosmology and Fundamental Physics \linebreak
  $\square$  Stars and Stellar Evolution \hspace*{1pt} $\square$ Resolved Stellar Populations and their Environments \hspace*{40pt} \linebreak
  $\square$    Galaxy Evolution   \hspace*{45pt} $\square$             Multi-Messenger Astronomy and Astrophysics \hspace*{65pt} \linebreak
  
\textbf{Principal Author:}

Name: Elena Gallo
 \linebreak						
Institution:  University of Michigan
 \linebreak
Email: egallo@umich.edu
 \linebreak
 
\textbf{Co-authors:}  \linebreak
Edmund Hodges-Kluck (NASA Goddard Space Flight Center)
 \linebreak
   Tommaso Treu (University of California, Los Angeles)
   \linebreak
 Jenny Greene (Princeton University)
  \linebreak
  Belinda Wilkes (Smithsonian Astrophysical Observatory)
     \linebreak
       Anil Seth (University of Utah)
         \linebreak 
  Amy Reines (Montana State University)
    \linebreak
    Vivienne Baldassare (Yale University)
  \linebreak
  Richard Plotkin (University of Nevada)
   \linebreak
    Rupali Chandar (University of Toledo)
     \linebreak

\justify

\textbf{Abstract:}  This document illustrates the feasibility of a few per cent level measurement of the local black hole occupation fraction in low mass galaxies through wide-field, high angular resolution X-ray imaging observations of local volume galaxies. The occupation fraction, particularly at the low end of the galaxy luminosity function, is a key benchmark for any model which aims to reproduce the formation and growth of super-massive black holes and their host galaxies. Our proposed measurement will complement orthogonal efforts that are planned in X-rays at high red-shifts, as well as in the local Universe with ground-based facilities. 

\pagebreak

\setcounter{page}{1}
\pagenumbering{arabic}

\section{The local black hole occupation fraction in a cosmological context}
Two quantities necessary to understand the connection between massive black holes (BHs) and their host galaxies are the active fraction (\factive) and occupation fraction (\focc) in the present-day universe, defined respectively as the fraction of galaxies with an AGN (down to an arbitrary luminosity) and the fraction of galaxies with a central BH, regardless of its activity level. Knowledge of \focc\ is also key to completing the inventory of BHs across the mass spectrum, and thus establishing the black hole mass density function.  For massive galaxies, dynamical measurements indicate that \focc\ is of order $\sim$1. However, owing to their smaller sphere of influence and lower accretion powered luminosity, \focc\ becomes increasingly hard to measure moving down the mass function, particularly in dwarf galaxies. \textit{This document illustrates the feasibility of a few per cent level measurement of the local BH occupation fraction in low mass hosts through wide field, high angular resolution X-ray imaging observations of local volume galaxies.}\\

To calibrate their prescriptions for BH growth, semi-analytical and numerical models alike rely on {(i)} the measured luminosity function of AGN and quasars, combined with {(ii)} the inferred BH mass density. The latter is estimated from the $z$-dependent galaxy luminosity function, folded with {(iii)} the local stellar velocity dispersion:BH mass relation, under the assumption that \textit{all} $z$=0 galaxies host a central, massive BH\cite{natarajan14,sijacki15,pacucci17}. Despite strong evidence for their ubiquity at the high mass end of the galaxy population, evidence for massive BHs is scarce, and hard to acquire\cite{ng18,ng19}, below host stellar masses log(\mstar/\msun)$\simlt $10, where the BH:host galaxy scaling relations also exhibit large scatter\cite{reinescomastri,kormendy16}.

\begin{figure}
\begin{center}
\vspace{-1.15cm}
\includegraphics[width=0.8\textwidth]{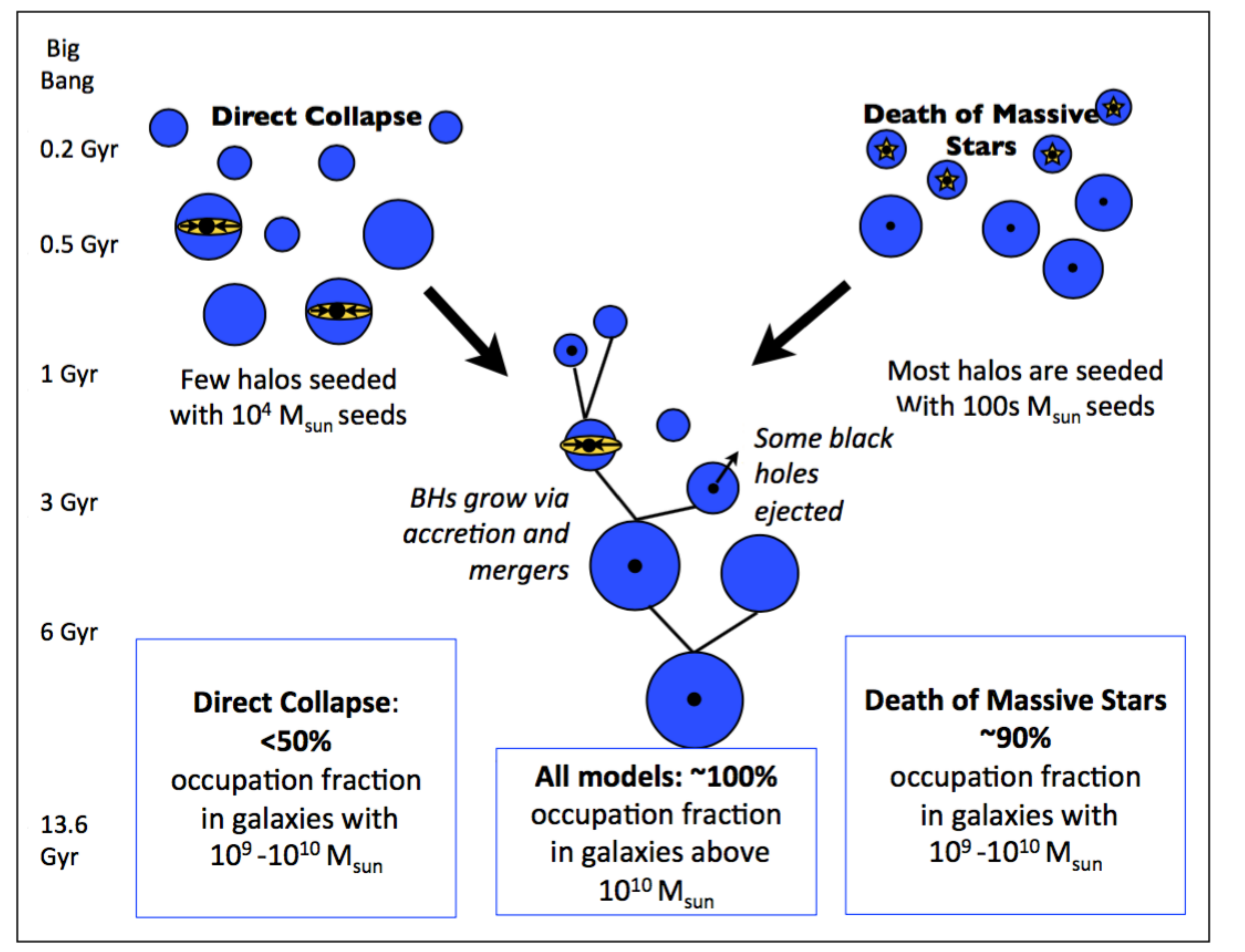}
\end{center}
\caption
{Qualitative predictions for the local BH occupation fraction from semi-analytical models\cite{Greene12}. Low mass galaxies are expected to have retained memory of their predominant BH seeding mechanism. 
\label{fig:seeds}
}
\end{figure}

In addition to providing a much needed $z$=0 anchor for the BH mass function, quantifying the BH occupation fraction in this regime is crucial to a number of high-stakes astrophysical problems. Between 9$\simlt$log(\mstar/\msun)$\simlt$10, the fraction of galaxies hosting BHs is thought to depend sensitively on the predominant mechanism by which BHs were `seeded' at high redshifts, with semi-analytical models suggesting that, by $z$=0, low mass  galaxies are much more likely to contain massive BHs if low-mass, Pop III remnants provide the \textit{dominant} seeding mode, as opposed to global gas collapse on galaxy-size scales, leading to larger but rarer seeds\cite{volonterinatarajan09,volonteri10,volonteri12,natarajan14,agarwal16,valiante16,ricarte18}  (Fig. \ref{fig:seeds}). Additionally, {\it if} the occupation fraction of dwarf galaxies were close to 100\%, then their BHs (rather than supernovae) could be entirely responsible for quenching star formation in these systems\cite{silk18,dickey19}. Finally, the expected rate of tidal disruption events is hugely sensitive to the BH occupation fraction at the low mass end of the galaxy population\cite{stonemetzger}.
%

\section{Measuring \focc\ from high resolution X-ray imaging}
\label{section:x-ray}

\subsection{  Methods}

The number of detected, actively accreting BHs in a sample of galaxies \textit{that is unbiased with respect to nuclear properties} constrains \focc. The most efficient way to detect such BHs down to the lowest Eddington ratios is through X-ray observations, as major issues/contaminants such as absorption by dust, incompleteness due to star formation/host contamination, and unknown massive star physics are all mitigated, if not totally eliminated, by working at X-ray energies. The basis of the constraint on \focc\ is an empirical relationship between the BH X-ray luminosity, \lx, and the stellar mass of its host galaxy, \mstar, that makes \factive\ a function of the sensitivity (see Fig.~\ref{figure.ngal_focc}). This relation likely results from the \mbh:\mstar\ relation and the small amount of fuel needed to sustain weak X-ray activity. If \focc$=$1, then a survey with arbitrary sensitivity would detect a nuclear X-ray source in every galaxy. The ratio between the \textit{expected} and \textit{observed} number of sources thus constrains \focc. 

\begin{figure}[tp]
\centering
\includegraphics[height=7cm]{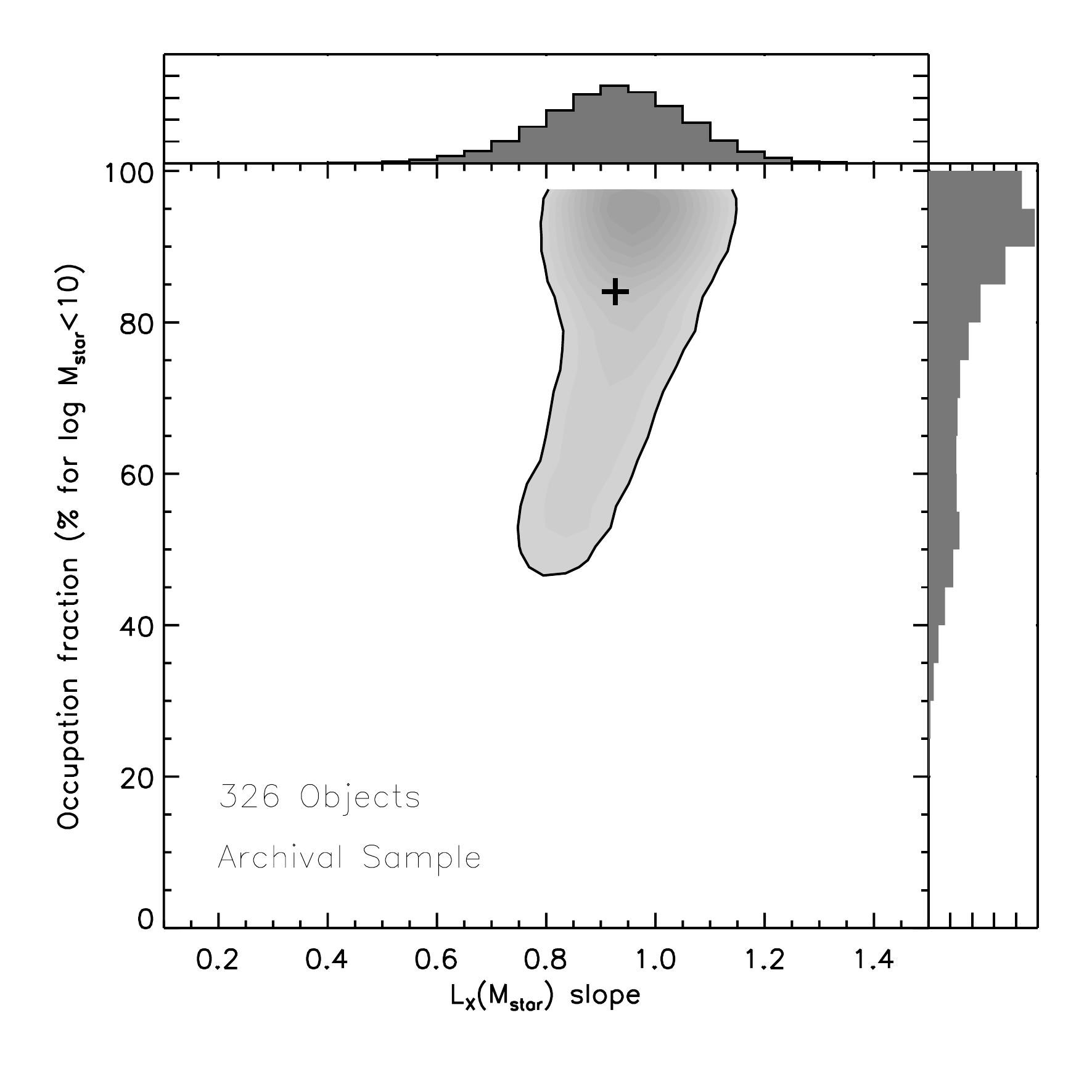}
\includegraphics[height=8cm]{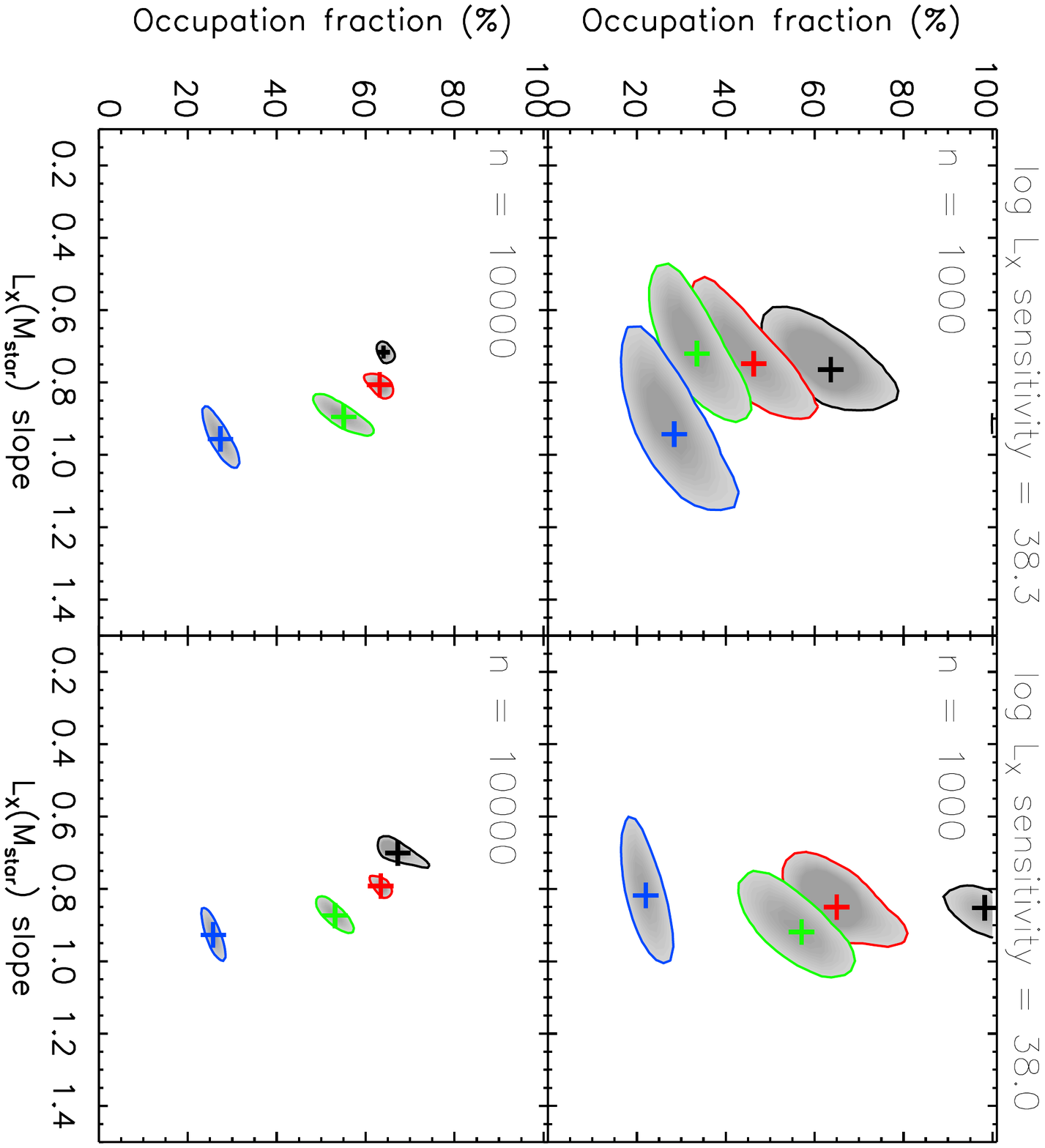}
\caption{
\textit{Left}: The best constraints (68\% C.L.) on \focc\ ($>40$\% for \mstar$\lesssim 10^{10}$\msun) and the $L_X$:\mstar\ relationship from a Bayesian analysis of \Chandra\ observations of 326~galaxies\cite{Miller15}.
\textit{Right}: These constraints are improved with more galaxies, as shown by the \focc\ and $L_X$:\mstar\ recovered for 1,000 or 10,000 simulated galaxies, based on a plausible catalog\cite{Blanton05}, at two sensitivities and with different input \focc. Using a \mstar-dependent sensitivity reduces $N_{\text{gal}}$ to $\sim$3,000 galaxies for 5\% precision.  
}
\label{fig:now}
\end{figure}

Prior surveys\cite{Gallo08,zhang09,Gallo10,Miller12,she17}, with the \cxo\ X-ray Telescope, probe Eddington ratios as low as $\sim 10^{-8}-10^{-5}$ (whereas \textit{bona fide} AGN have $\simgt 10^{-3}$\cite{ho08}).
The best current constraint below $\log$(\mstar/\msun)$\le 10$ comes from \cxo\ data for 326 galaxies within 20~Mpc, finding that \focc~$>$~40\% (68\% C.L.; Fig.~\ref{fig:now}, let panel). This approach must account for contamination by high Eddington ratio X-ray binaries (XRBs), which have similar \lx\ to weakly accreting BHs. This is difficult to do on a case-by-case basis, but a statistical assessment is possible\cite{foord17,Lee19}. To first order, the expected luminosity (\lxrb) can be related to \mstar: for low-mass XRBs \lxrb~$\propto$~\mstar, whereas for high-mass XRBs \lxrb~$\propto$~SFR (the star-formation rate)\cite{Lehmer10,Mineo12,Lehmer16}. Along the ``main sequence'' of star-forming galaxies\cite{Brinchmann04,Pearson18}, SFR~$\propto$~\mstar$^{\alpha}$. Since XRB formation is (to first order) a local process, the nuclear \lxrb\ is related to the stellar mass in the nucleus. XRBs are discrete and follow a near-universal luminosity function\cite{Gilfanov04,Mineo12}, so the \lxrb\ can be converted into the probability of detecting a bright nuclear XRB or a combination of unresolved XRBs, \pxrb. The contamination in existing surveys is several percent\cite{Gallo08,Gallo10,Miller12,Miller15,foord17,Lee19}.

The combination of \pxrb\ and the expected BH detection rate (\factive) implies a number of galaxies needed to constrain \focc\ to some precision, $N_{\text{gal}}$. For example, detecting only 50/300 sources when the combination of \pxrb\ and \factive\ implies an expected detection rate of 100/300 rules out \focc$=$1 with high confidence. In this simplified picture, Fig.~\ref{figure.ngal_focc} shows the number of galaxies ($N_{\text{gal}}$) needed to constrain \focc\ to 5\% precision (68\% confidence interval) as a function of \pxrb\ and \factive, at a given \mstar. Clearly, this is most feasible for \factive~$\gtrsim$~0.3 and \pxrb$<0.1$ (the bottom right corner), and simulations assuming these values and based on a realistic distribution of galaxy masses\cite{Blanton05} (\ref{fig:now}, right panel) indicate that \textit{about 3000~galaxies are needed to measure \focc\ between $8 < \log$~\mstar~$<10$ in mass bins of 0.5~dex.}

\subsection{Instrumental requirements}

\begin{figure}
\vspace{-6.5cm}
\includegraphics[width=0.85\textwidth]{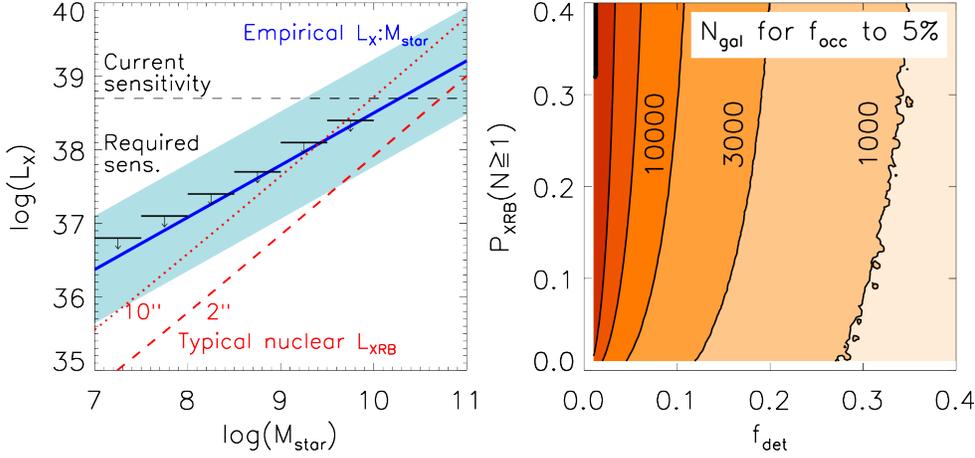}
\vspace{-5.6cm}
\caption{
\textit{Left}: A schematic diagram of the $L_X$:\mstar\ relation (blue line) that enables the measurement of \focc: the relationship predicts a detection rate at a given sensitivity (black lines), and the \textbf{ratio of observed to expected number constrains \focc}. In practice, one simultaneously constrains $L_X$:\mstar\ and \focc\cite{Miller12,Miller15}. The dashed and dotted red lines show the average nuclear \lxrb\ for a galaxy at $d=50$~Mpc for two detection cell sizes. \lxrb, combined with the XRB luminosity function\cite{Gilfanov04,Mineo12}, yields the likelihood (\pxrb) of detecting an XRB. High resolution helps to reject off-nuclear XRBs. 
\textit{Right}: The number of galaxies needed to measure \focc\ to 5\% precision as a function of \factive\ and \pxrb. This precision is feasible with an instrument and sample that reaches the lower right corner.
}
\label{figure.ngal_focc}
\end{figure}

To maximize \factive, minimize \pxrb, and observe a large number of galaxies requires:   \smallskip\\
%
\textbf{Collecting area---Maximizing \factive:} $N_{\text{gal}}$ is a strong function of \factive, which depends on the sensitivity. Fig.~\ref{figure.ngal_focc} shows that $\sim$100x better sensitivity than existing surveys is needed to constrain  \factive~$\gtrsim$~0.3 in the low mass (dwarf) regime where \focc\ is sensitive to the predominant BH seeding mechanism (i.e. between $7<\log$\mstar$<10$). Observing enough galaxies within a nominal mission lifetime (3-5~years) requires at least \textit{10~times greater collecting area than \cxo}.    \smallskip\\
\textbf{High resolution---Minimizing \pxrb:} Sharp images enable the rejection of off-nuclear X-ray sources and reduce the chance of confusing flux from multiple unresolved XRBs with a nuclear source. We simulated observations based on realistic galaxies and X-ray facilities and found that \pxrb$\propto \theta$. \textit{A half-power diameter $\simlt $0.6~arcsec is needed to maintain \pxrb$<0.1$ for galaxies within a distance of $\sim$200~Mpc.}    \smallskip\\
%
\textbf{Wide field of view---Serendipitous sources:} The \Chandra\ resolution degrades sharply off-axis and the FOV with sub-arcsec resolution is only a few arcmin$^2$. \textit{Widening the FOV within which $\theta < 0.6$~arcsec to at least $15\times 15$~arcmin dramatically improves the observing efficiency}, as many more serendipitous galaxies will be in-field, and fewer exposures are needed to tile dense regions.  \smallskip\\

\textbf{These criteria are met by two mission concepts: the \textit{Lynx} X-ray Surveyor\cite{Gaskin18} and the Advanced X-ray Imaging Satellite\cite{mushotzky18} (\textit{AXIS}) Probe}. The \textit{Lynx} High Definition X-ray Imager has an area of 20,000~cm$^2$ at 1~keV, an on-axis half-power diameter$<$0.5~arcsec, and $<$1~arcsec across the $23\times 23$~arcmin FOV. \textit{AXIS} has a smaller collecting area of 7,000~cm$^2$ at 1~keV, but a similar half-power diameter and FOV. 
\subsection{  Notional observing strategy}

Based on the \textit{Lynx} and \textit{AXIS} science drivers and notional exposure times, $N_{\text{gal}}$ can be accumulated through commensal surveys and serendipitous sources at $\log$~\mstar~$>8.5$, from observations of cluster outskirts, deep fields, and large galaxies, while targeted snapshots are needed at lower masses:    \smallskip\\
\textit{Galaxy cluster outskirts:} Mosaics of cluster outskirts to study accreting gas and plasma physics require long exposures. The combination of relatively high galaxy density and low intra-cluster medium surface brightness makes these fields sensitive probes of \focc\ at $\log$~\mstar~$\gtrsim$8.5. Based on optical maps, complete mosaics of nearby clusters will contain $\sim$2,000 targets (total).   \smallskip\\ 
\textit{Deep fields:} Surveys to count high-redshift AGNs will surpass the 4~Ms \Chandra\ Deep Field-South, which already probes AGNs in sub-$L^*$ galaxies in a cosmological volume. Assuming a uniform Eddington ratio distribution\cite{aird12}, the distribution of X-ray sources probes \focc\cite{Miller15}, especially at $\log$\mstar$\ge 10$, and places tight constraints on the slope and scatter in the $L_X$:\mstar\ relationship.   \smallskip\\
\textit{Normal galaxies:} Many science goals for the concept missions target massive galaxies ($\log$~\mstar~$>10$). Since dwarfs are clustered around massive galaxies\cite{binggeli90}, such observations will yield useful, low-mass targets. Many will be unsuitable due to irregular morphology or high background, but 50-100~ks observations of galaxies within 100~Mpc will build up a sample of 2,000-6,000 dwarfs with $\log$~\mstar~$<9$, based on the galaxy luminosity function\cite{schechter76} and the \textit{Lynx} and \textit{AXIS} mirrors.   \smallskip\\
\textit{Targeted survey:} There will be too few serendipitous galaxies with $\log$~\mstar~$<8.5$ because typical observations are only sensitive to these BHs within 25~Mpc (i.e., detecting such a BH at a larger distance is rare). This motivates a snapshot survey of 200-400 nearby dwarfs with exposures of 5-15~ks. The total observing time can be reduced to $\sim$1.5~Ms (a very large \cxo\ program) by selecting fields with multiple galaxies, such as in Virgo.

{\section{Synergistic Efforts}}

From a conceptual standpoint, the same investigation as described above can be carried out at different wavelengths, provided that a comparably efficient diagnostics is identified for detecting highly sub-Eddington, massive BHs. Whereas traditional, optical emission line ratio diagnostics is only moderately efficient at detecting low Eddington ratio BH activity (see, however, \cite{trump15}), flat spectrum, compact, nuclear radio sources have the potential to reveal a large population of BHs in local low mass galaxies. Moreover, the radio to X-ray luminosity ratio of (radiatively inefficient) accreting BHs is a rising function of BH mass, through the so-called Fundamental Plane (FP) of BH activity\cite{merloni03,falcke04}. Using the FP as a reliable mass estimator, however, requires simultaneous multi-wavelength coverage as well as high quality radio and X-ray data (this is again for the purpose of disentangling low Eddington ratio accretion onto a massive, central BH from $\sim$Eddington limited XRBs). This is best achieved by \cxo-like quality X-ray imaging, in concert with large baseline radio interferometry. Specifically, the development of the {next generation VLA} would dramatically increase the search volume, pushing the search for radio emission from sub-Eddington BHs to about 1 Gpc  \cite{plotkinreines} (to this end, it is important to notice that, owing to the large scatter to the FP, upper limits are not nearly as informative for the purpose of discriminating between massive BHs and bright XRBs\cite{gultekin19}).

In the X-ray regime, orthogonal efforts to identify the seeds of today's billion \msun\ BHs are being planned in the high-redshift domain, where wide-field, high resolution X-ray imaging can yield \textit{direct} detection of $\simlt 10^5$ \msun\ BHs up to $z=$10, and answer the question whether heavy seeds ever form (regardless of whether they are the dominant seeding mechanism; see WP by Haiman et al.).

Lastly, with a factor $\sim$5 increase in angular resolution over current ground-based facilities, Extremely Large Telescopes will be sensitive to \textit{dynamical} signatures of tens $10^3$ \msun\ BHs in low mass galaxies at the distance of Andromeda -- and $10^4$ \msun\ BHs out to 5 Mpc. These targeted studies will answer complementary questions about BH feedback at low mass end of the mass spectrum, e.g. by establishing whether local empirical scaling relations between BH mass and host galaxy properties extend to the dwarf regime or exhibit substantial deviations (see WP by Greene et al.).

\clearpage
\section{References}

\renewcommand{\section}[2]{}
{  
\bibliographystyle{nature}

}

\end{document}